\documentclass[prl,aps,showpacs,twocolumn]{revtex4}
\usepackage{graphicx}

\begin{document}

\preprint{}

\title{Chromatin dynamics: Nucleosomes go mobile through twist defects}% Force line breaks with \\

\author{I. M. Kuli\'{c} and H. Schiessel}
\affiliation{%
Max-Planck-Institut f\"{u}r Polymerforschung, Theory Group, POBox
3148, D 55021 Mainz, Germany}

\date{\today}

\begin{abstract}
We study the spontaneous ''sliding'' of histone spools
(nucleosomes) along DNA as a result of thermally activated single
base pair twist defects. To this end we map the system onto a
suitably extended Frenkel-Kontorova model. Combining results from
several recent experiments we are able to estimate the nucleosome
mobility without adjustable parameters. Our model shows also how
the local mobility is intimately linked to the underlying base
pair sequence.
\end{abstract}

\maketitle

The genetic information of all higher organisms is organized in
huge beads-on-a-chain arrays consisting of centimeters to meters
of DNA wrapped around globular aggregates of so-called histone
proteins. The basic unit of chromatin, the nucleosome, is a tiny
$10\times 5\times 6$~nm sized spool composed of $147$ base pairs
(bps) DNA tightly wrapped around an octamer made from 8 histone
monomers. Each nucleosome is connected via a stretch of ''linker''
DNA to the next such protein spool. The wrapped DNA, being coiled
in $\sim 1 \frac{3}{4}$ turns of a left handed helix with radius
$\sim 4.2$~nm, is strongly distorted from its preferred straight
ground state due to strong interactions with the histone octamer,
namely short range electrostatics (between the negatively charged
DNA sugar-phosphate backbone and the positively charged octamer
surface) and through extensive hydrogen-bonding -- both localized
at 14 discrete interaction patches helically arranged along the
octamer surface~\cite{luger97}.

Higher order structures, from the 30nm-chromatin fiber up to the
highest level of DNA condensation, the fully folded chromosome,
are designed to achieve a huge DNA volume fraction. They all rely
on the significant stability of the nucleosome complex. On the
other hand, fundamental life processes like transcription (making
RNA offprints from the underlying DNA) and DNA replication seem to
be in conflict with the picture of a stable nucleosome, as they
are all performed by protein machines that track the DNA helix.
The latter inevitably implies that every DNA bound obstacle
(protein) has to be penetrated or even completely removed from its
DNA target. In fact, the numbers are quite dramatic: A typical
gene extends over hundreds of nucleosomes, each contributing
$30-40$ $k_{B}T$ net adsorption
energy~\cite{polach95,brower-toland02}. Also other mechanisms like
the activation of genes rely on regulatory protein binding to
specific DNA sequences that are often covered by nucleosomes
making them inaccessible.

A key to the understanding of these seemingly contradictory
features might be the {\it physical} phenomenon of thermally
driven nucleosome ''sliding'' along DNA (also called nucleosome
repositioning) which has repeatedly observed in well-defined {\it
in vitro}
experiments~\cite{beard78spadafora79,pennings91,flaus98gottesfeld02},
reviewed in Ref.~\cite{schiessel03b}. Spontaneous repositioning is
strongly temperature dependent; at room temperature nucleosomes
move a few tens of bps within an hour~\cite{pennings91}. Despite
clear evidence for repositioning the underlying mechanism has been
the matter of longstanding controversy, especially due to the lack
of any quantitative theoretical treatment of nucleosome statics
and dynamics that has to rely on the detailed knowledge of the
molecular structure and its underlying parameters.

Only very recently -- since the documentation of the high
resolution X-ray structure~\cite{luger97} and the presentation of
other new
experiments~\cite{polach95,brower-toland02,flaus98gottesfeld02} --
this has become possible. First theoretical models of nucleosome
repositioning~\cite{schiessel01c,kulic02} assume that it is based
on the formation of DNA loop defects that form on either end of
the nucleosomal DNA followed by their thermal diffusion around the
octamer, similar to the de Gennes-Edwards reptation mechanism.
This model seems to be successful in explaining the apparent 10
bps quantization of the nucleosome ''jump''
length~\cite{pennings91} and it also reproduces the observed
diffusion constants. Recent
experiments~\cite{flaus98gottesfeld02}, however, indicate a more
local 1 bp-step mechanism that cannot be understood within this
model. This lead us here to consider an alternative mechanism:
twist diffusion. The carrier of motion in this case is a twist
defect that contains one missing or one extra bp.

The possibility of twist defects was demonstrated as soon as the
high resolution crystal structure of the core particle (the
octamer plus wrapped DNA) was resolved~\cite{luger97}. In that
study the core particles were reconstituted from palindromic 146
bp DNA and histones assuming that this would result in a complex
with perfect two-fold symmetry. However, it turned out that one bp
is localized at the dyad axis, the rest being divided into a 73 bp
half and a 72 bp half. The missing bp of the shorter half is,
however, not localized at its terminus but instead at a 10 bp
stretch close to the dyad axis (cf.~Fig.~4d in
Ref.~\cite{luger97}). This is due to the attraction between the
DNA termini of adjacent particles in the crystal that come close
to mimic a bp step at the cost of forming a twist defect far
inside the wrapped chain portion. This allows us to estimate the
energy for a single defect to be smaller than the stacking energy
that is $\sim 10-20k_{B}T$~\cite{estimate}.

In order to model the twist diffusion mechanism we map the
nucleosomal DNA on a Frenkel-Kontorova (FK) chain of particles
connected by harmonic springs in a spatially periodic potential
(cf.~Fig.~1). The original FK model was introduced more than sixty
years ago to describe the motion of dislocations in
crystals~\cite{frenkel38}. In the meantime variants of this model
were applied to many different problems including charge density
waves~\cite{floria96}, sliding friction~\cite{braun97,strunz98},
ionic conductors~\cite{pietronero81,aubry83}, chains of coupled
Josephson junctions~\cite{watanabe96} and adsorbed atomic
monolayers~\cite{uhler88,schiessel03}. Here, in the context of DNA
adsorbed on the octamer, the beads represent the base pairs. The
springs in between have an equilibrium distance $b=0.34$ nm and a
constant that reflects the coupled DNA twist-stretch elasticity.
Specifically
\begin{equation}
E_{elastic}\left( \left\{ x_{n} \right\} \right) =\sum_{k}C\left( \frac{%
x_{k+1}-x_{k}}{b}-1\right) ^{2}  \label{eelastic}
\end{equation}
Here the conformation of the wrapped DNA is given by the set
$\left\{ x_{n} \right\}$ where $x_{n}$ is the position of the
$n$th bp measured along the helical backbone; $C\simeq
70-100k_{B}T$ is the combined twist and stretch spring constant
including the (here unfavorable) twist-stretch
coupling~\cite{kamien97marko97} and the summation goes over all bp
associated with the wrapped DNA. In addition there is the external
potential of the 14 contact points to the octamer with neighboring
points being 10 bp apart~\cite{luger97} that we model as follows
\begin{eqnarray}
E_{ads}\left( \left\{ x_{n} \right\} \right)
&=&-U_{0}\sum_{k}\sum_{l=1}^{14} \left( \left( \frac{x_k
-10bl}{a}\right) ^{2}-1\right) ^{2} \nonumber \\
&&\times \theta \left( a-\left| x_{k}-10bl\right| \right)
\label{eads}
\end{eqnarray}
with $\theta $ being the Heaviside step function. The
two parameters of the external potential, its depth $U_{0}$ and its width $a$%
, can be estimated as follows. $U_{0}$ represents the pure
adsorption energy per point contact which follows from competitive
protein binding~\cite{polach95} to be of order $6k_{B}T$. The
other parameter, $a$, can be estimated from the fluctuations of
the DNA in the crystal measured by the B-factor (cf.~Fig.~1b
in~\cite{luger97}) at different
nucleosome positions. The ratio of DNA helix fluctuations $%
R_{fluct}=\left\langle x_{middle}^{2}\right\rangle /\left\langle
x_{bond}^{2}\right\rangle \approx 3$ at positions between the
binding sites and at the bound sites is a measure of DNA
localization. Using a quadratic expansion of Eq.~\ref{eads} one
finds from a straightforward normal mode
analysis that $a=\left( 5U_{0}/[\left( R_{fluct}-1\right) C]\right) ^{1/2}%
b\sim b/2$, i.e., the adsorption regions lead to a strong
localization of the DNA. Knowing all involved parameters the total
energy of the DNA chain confined in the nucleosome can be written
down
\begin{equation}
E_{tot}=E_{elastic}+E_{ads}+E_{sd} \label{etot}
\end{equation}
The last term $E_{sd}$ is the sequence dependent part of the total
energy which we will neglect first. In the following we study the
mechanism for thermal motion of DNA governed by $E_{tot}$.
Generally two scenarios are possible: ({\it i}) The generation of
kink-antikink pairs inside the nucleosome and ({\it ii}) a kink
(or antikink) injection at either nucleosome end. Since the first
mechanism is energetically roughly twice as costly than the second
one, we will focus here on the (anti)kink injection mechanism
only.

\begin{figure}
\includegraphics*[width=8cm]{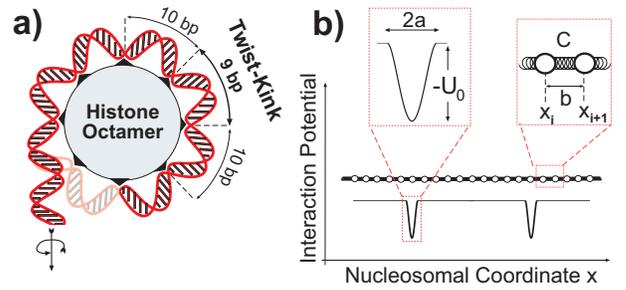}
\caption{The twist-diffusion mechanism for nucleosome
repositioning. a) A concerted translational and rotational motion
of DNA leads to injection of twist-defects (kinks) which migrate
between the octamer adsorption sites (black triangles) leading to
a ''creep'' motion of DNA. b) The corresponding Frenkel-Kontorova
model for twist diffusion and its characteristic parameters (cf.
text for details).} \label{f.1}
\end{figure}

How and how fast does the kink step around the nucleosome? Due to
the strong DNA localization at the binding sites ($a/b<1$) for a
realistic range of parameters $U_{0}$ and $C$ the kink is
localized either between two adsorption positions, i.e., smeared
out over $10$ bp (denoted by the $K_{10}$ state), or between three
of them, i.e., smeared out over $20$ bp (the $K_{20}$ state). It
is obvious that the motion of a (anti)kink will consist of an
alternation between $K_{10}$ and $K_{20}$ states similarly to an
earthworm creep motion. To model this process we introduce the
effective kink coordinate $x_{K}$ describing the coordinate of the
DNA bp being
pinned/depinned during a single kink step, so that $x_{K}\approx 0$ and $%
x_{K}\approx b/2$ correspond to $K_{10}$ and $K_{20}$,
respectively, whereas $x_{K}\approx b$ means that the kink moved
by one bp step. The Peierls-Nabarro potential experienced by the
kink is then given by $U_{PN}\left( x_{K}\right) = C_{{\it
eff}}\left( x_{K}/b-1/2\right) ^{2}-U_{0}\left(
x_{K}^{2}/a^{2}-1\right) ^{2}$ for $0<x_{K}<b/2$ and
$U_{PN}\left( x_{K}\right) = U_{PN}\left( b-x_{K}\right)$ for $b/2\leq x_{K}<%
b$. Here $C_{{\it eff}}=\frac{2}{10\pm 1}C$ with the ''$-$'' sign
referring to a kink (1 bp missing) and the ''$+$''sign to an
antikink (1 additional bp). Depending on the ratio of parameters
$U_{0}$ and $C$, the state $K_{20}$ corresponds to a local minimum
or maximum of $U_{PN}$ whereas $K_{10}$ is always stable for the
relevant parameter range. The rate for the kink step process is
then given by the expression $f_{step}=k_{B}Tj_{0}/b^2\zeta _{{\it
eff}}$
with $j_{0}^{-1}=\left( \int_{0}^{1} e^{-U_{PN}\left( sb%
\right) /k_{B}T} ds\right) \left( \int_{0}^{1} e^{+U_{PN}\left(
sb\right) /k_{B}T} ds\right) $ and $\allowbreak \zeta
_{eff}=\frac{4\pi ^{2}}{10b}\mu _{spin}$, the effective kink
friction constant. Here $\mu _{spin}=1.3\times 10^{-20}N s$ is
roughly the rotational friction for a single
basestep~\cite{levinthal56nelson99}. To determine the rate at
which twist defects are formed at the entry/exit points of the DNA
one can now use an argument similar to the one presented in
Ref.~\cite{schiessel01c}: The ratio of the life time $t_{life}$ of
a kink to the time interval $t_{inj}$ between two kink injection
events at the end of the wrapped DNA portion equals the
probability to find a defect on the nucleosome, i.e.
$t_{life}/t_{inj}\simeq N_{site}e^{-U_{Kink}/k_{B}T}$. Here
$N_{site}=13$ denotes the number of possible positions of the
defect between the 14 binding sites and $U_{Kink}\simeq C/10 $ is
the energetic cost for a single kink (cf. above).

How is the average life time $t_{life}$ of a defect related to
$t_{step}$, the typical time needed for one step? This can be
determined from the mean first passage times $\tau _{left}$ and
$\tau_{right}$ for a defect that forms, say, at the left end to
leave the nucleosome at the same or at the other end,
respectively. From Ref.~\cite{vankampen} one finds $\tau
_{left}=\left( 25/6\right) t_{step}$ and $\tau
_{right}=28t_{step}$. Furthermore, the probability to leave at the
left end is $p_{left}=12/13$ and at the right end
$p_{right}=1/13$~\cite{vankampen} which gives the life
time as the weighted average $t_{life}=6t_{step}$. Only a fraction $%
p_{right} $ of the defects reaches the other end and will lead to
a repositioning step, i.e., the time of a 1bp diffusion step of
the nucleosome along the DNA is given by $T=t_{inj}/p_{right}$.
Putting all this together we arrive at $T\simeq 6b^2 \zeta_{{\it
eff}}j_{0}^{-1}/k_{B}T\exp \left(
C/10k_{B}T\right) $. For realistic parameter values $C=100k_{B}T$, $%
U_{0}=6k_{B}T$ and $R_{fluct}=3$ we find $T\simeq \allowbreak
10^{-3}$ s implying a nucleosome diffusion constant $D=580 \mbox{
bp}^{2}/\mbox{s}=$ $6.6\times 10^{-17}\mbox{ m}^{2}/\mbox{s}$.

\begin{figure}
\includegraphics*[width=8cm]{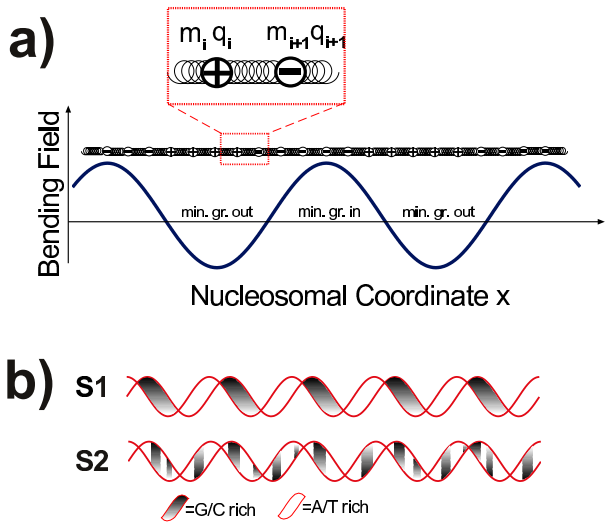}
\caption{The ''charged'' Frenkel-Kontorova extension of the model
in Fig.~1b) including effects from anisotropic sequences. a) In
addition to the contact points the DNA sequence couples linearly
to an octamer-fixed ''bending field'' through the anisotropic
bending parameters $q_{i}$ (''bending charge''). b) Two sequences
with extremely different mobilities. S1: highly anisotropic, 10 bp
phased (''TG''-like) sequence with $D_{sd}\approx
10^{-4}-10^{-5}\mbox{ bp}^{2}/\mbox{s}$. S2: random sequence
corresponding to $>95\%$ of the genome with $D_{sd}\approx
10^{2}\mbox{ bp}^{2}/\mbox{s}$.} \label{f.2}
\end{figure}

Hence we find repositioning rates that are orders of magnitude
faster than the ones observed in experiments~\cite{pennings91}.
Even worse, the experimental observation of an apparent 10 bp jump
length~\cite{pennings91} seems to be inconsistent with our
predictions. We show now how these facts can be explained by the
existence of additional barriers with a 10 bp periodicity. To do
so we have to extend our simple model to deal with the quenched
disorder stored in the DNA bp sequence. The sequence dependent
anisotropic bendability, i.e., the propensity of DNA to bend in
different directions with different elastic constants turns out to
be essential. It has been known for
long~\cite{satchwell,shrader89shrader90} that (A/T) rich
dinucleotide steps (dns) prefer to face the octamer in the minor
groove (i.e., at the octamer contact points) whereas (G/C) rich
dns prefer to face the octamer in the major groove (i.e., between
contact points). This reflects different propensities of the
dinucleotides to widen or compress towards the DNA minor groove.
To incorporate these anisotropic effects into our model we first
note that the bending state of the DNA molecule is fully
constrained by its helical path on the octamer surface. Moving a
DNA sequence via twist diffusion by a few bp ($<10$ bp) along that
path changes the relative rotational setting of the bent DNA with
respect to its preferred bending direction causing an energetic
penalty, whereas a motion by 10 bp restores the initial rotational
setting.
We address this by introducing a 10bp periodic ''bending field'' $%
F_{bend}\left( x\right) =-\cos \left[ 2\pi x/\left( 10b\right) %
\right] $ attached to the octamer surface. We assume the DNA
sequence to couple linearly to that field through ''bending
charges'' $q_{k}$ attached to each of the dns. This gives us
finally the third term in Eq.~\ref{etot}:
\begin{equation}
E_{sd}=\sum\nolimits_{k}q_{k}F_{bend}\left( x_{k}\right) +m_{k}
\label{E_seq_dep}
\end{equation}
In addition to the anisotropic term we also introduced here the
isotropic bending parameters $m_{k}$ to include isotropic
flexibility effects (which become important when the $q_{k}$'s
vanish or average out). The summation involved is again over all
base pairs~\cite{footnote1} incorporated in the nucleosome.
$q_{k}$ and $m_{k}$ both have units of energy and can be extracted
from competitive protein binding
experiments~\cite{shrader89shrader90} for each of the 10 dns (AA,
AT, GC...). To obtain a rough estimate we distribute the dns into
three classes: 1) (G/C) containing dns, 2) (A/T) containing dns
and 3) mixed dns (like AG, CT etc.) and treat the dns in each
class as identical. Using the available experimental
data~\cite{shrader89shrader90,Cacchione97} we then arrive at $%
q_{G/C}\approx $ $\ 95$, $q_{A/T}\approx -85$\ , $q_{mixed}\approx 0$ \ and $%
m_{G/C}\approx 20,$\ $m_{A/T}\approx -3$, $m_{mixed}\approx 7$,
where all energies are in cal/mol per dns.

It turns out that the nucleosome mobility depends strongly on the
underlying bp sequence. When shifting the position of all beads by
$l$ bp steps, $x_{k}\rightarrow $ $x_{k}+lb$, we find
$E_{sd}\left( l\right) =\left(A/2\right)\cos \left( 2\pi l/10-\phi
\right) $ to vary as a cosine function of $l$ with phase $\phi $
and amplitude $A$ determined by the DNA sequence, which is assumed
to be appropriately periodic here. Arranging G/C and A/T tracts
properly and taking the sequence dependent $q$ and $m$ values
given above we can easily reach amplitudes $A$ (i.e. barriers to
repositioning) that exceed $10-12$ kcal/mol! A very effective
sequence arrangement called the ''TG''-sequence which leads to a
strong nucleosome stability and localization was experimentally
constructed in Ref.~\cite{shrader89shrader90} by putting G/C
tracts around positions $k=0,10,20...$ and A/T tracts around
$k=5,15,25...$. In our picture this means to put the ''bending
charges'' $q$ along the DNA such that they couple favorably to the
bending field $F_{bend}$ for a distinct rotational setting whereas
a 5 bp shift is extremely costly (cf. Fig.~2). The 5S-RNA sequence
which was used in most nucleosome mobility experiments shows also
the effect of an optimal rotational setting. It is less pronounced
than in the ''TG'' case, yet it is still detectable. More involved
theoretical computations relying on molecular sequence dependent
deformability parameters~\cite{Mattei02} reveal barriers $%
A\approx 5-6$ kcal/mol for this particular sequence. The sequence
dependent barrier height $A$ exponentially suppresses the bare
(sequence independent) diffusion constant $D$ obtained above
leading to the sequence dependent diffusion constant $D_{sd}$:
\begin{equation}
D_{sd}=DI_{0}^{-2}\left( A/2k_{B}T\right) \approx \frac{\pi j_{0}A}{%
12\zeta_{{\it eff}}}e^{-\left( A+C/10\right) /k_{B}T}
\label{D_seq_dep}
\end{equation}
with $I_{0}$ being the modified Bessel function.

Equation \ref{D_seq_dep} predicts that mobility experiments with
highly anisotropic sequences like ''TG'' (instead of the standard
''5S-RNA'') would find hardly any appreciable repositioning on the
one hour timescale
{\it if} it would be solely mediated via twist defects ($%
D_{sd}=10^{-6}-10^{-7}\times D=10^{-4}-10^{-5}\mbox{
bp}^{2}/\mbox{s}$). The typical path for a nucleosome to escape
from such a rotational trap goes very likely via the previously
considered loop formation mechanism~\cite{schiessel01c,kulic02}
that allows ''tunneling'' over sequence barriers, thus dominating
over twist-diffusion for extremely anisotropic sequences. An
experimental test for this prediction would be to increase the
free DNA segment length which in this regime should strongly
enhance the loop mediated mobility~\cite{kulic02} whereas it would
leave the twist diffusion unaffected. Going to the other extreme,
in the most relevant case of random isotropically bendable
sequences which make up more than 95\% of the eucaryotic genome
one should observe that the twist diffusion mechanism is strongly
enhanced by 2-3 orders of magnitude as compared to the {\it in
vitro} measurements on ''5S-RNA''.

In conclusion the following picture is implied: On physiological
timescales the majority of genomic nucleosomes seems to be
intrinsically highly mobile. However, only a small fraction
($<5\%$) of all nucleosomes has strongly reduced mobility due to
anisotropic DNA sequences which they populate. We speculate that
only the latter require the action of active (ATP consuming)
remodelling mechanisms~\cite{lorch99} making them hotspots and
switching elements for global chromatin rearrangements.

We thank R. Bruinsma, K. Kremer, K. Luger, F. M\"{u}ller-Plathe
and J. Widom for helpful discussions.


\begin{thebibliography}{0}

\bibitem{luger97}  K. Luger, A. W. M\"{a}der, R. K. Richmond, D. F. Sargent, and
T. J. Richmond, Nature (London) \textbf{389}, 251 (1997).

\bibitem{polach95}  K. J. Polach and J. Widom, J. Mol. Biol. \textbf{254}, 130
(1995); \textbf{258}, 800 (1996).

\bibitem{brower-toland02}  B. D. Brower-Toland, C. L. Smith, R. C. Yeh, J. T. Lis,
C. L. Peterson, and M. D. Wang, Proc. Natl. Acad. Sci. USA
\textbf{99}, 1960 (2002).


\bibitem{beard78spadafora79} P. Beard, Cell \textbf{15}, 955 (1978); C. Spadafora, P. Oudet, and P. Chambon Eur. J. Biochem.
\textbf{100}, 225 (1979).

\bibitem{pennings91}  S. Pennings, G. Meersseman, and E. M.
Bradbury, J. Mol. Biol. \textbf{220}, 101 (1991); Proc. Natl.
Acad. Sci. USA \textbf{91}, 10275 (1994); G. Meersseman, S.
Pennings, and E. M. Bradbury, EMBO J. \textbf{11}, 2951 (1992).

\bibitem{flaus98gottesfeld02}  A. Flaus and T. J. Richmond,  J. Mol. Biol. \textbf{275}, 427
(1998); J. M. Gottesfeld, J. M. Belitsky, C. Melander, P. B.
Dervan, and K. Luger, J. Mol. Biol. \textbf{321}, 249 (2002).

\bibitem{estimate} R. L. Ornstein, R. Rein, D. L. Breen and R. D. Macelroy, Biopolymers
\textbf{17}, 2341 (1978)

\bibitem{schiessel03b} H. Schiessel, in preparation

\bibitem{schiessel01c}  H. Schiessel, J. Widom, R. F. Bruinsma, and W. M.
Gelbart, Phys. Rev. Lett. \textbf{86}, 4414 (2001); \textbf{88},
129902 (2002).

\bibitem{kulic02}  I. M. Kuli\'{c} and H. Schiessel, Biophys. J. (in
press).

\bibitem{frenkel38}  J. Frenkel and T. Kontorova, Zh. Eksp. Teor.
Fiz. \textbf{8}, 1340 (1938); 1939 \textit{J. Phys. (Moscow)
}\textbf{1}, 137 (1939).

\bibitem{floria96}  L. M. Floria and J. J. Mazo, Adv. Phys. \textbf{%
45}, 505 (1996).

\bibitem{braun97}  O. M. Braun, T. Dauxois, M. V. Paliy, and M.
Peyrard, Phys. Rev. Lett. \textbf{78}, 1295 (1997).

\bibitem{strunz98}  T. Strunz and F.-J. Elmer, \textit{Phys. Rev. E }%
\textbf{58}, 1601 (1998).

\bibitem{pietronero81}  L. Pietronero, W. R. Schneider, and S.
Str\"{a}ssler, Phys. Rev. B \textbf{24}, 2187 (1981).

\bibitem{aubry83}  S. Aubry, J. Phys. (France) \textbf{44}, 147 (1983).

\bibitem{watanabe96}  S. Watanabe, H. S. J. van der Zant, S. H. Strogatz, and
T. P. Orlando, Physica D \textbf{97}, 429 (1996).

\bibitem{uhler88}  W. Uhler and R. Schilling, Phys. Rev. B \textbf{37}, 5787 (1988).

\bibitem{schiessel03}  H. Schiessel, G. Oshanin, A. M. Cazabat and, M.
Moreau, Phys. Rev. E \textbf{66}, 056130 (2002).

\bibitem{kamien97marko97} R. D. Kamien, T. V. Lubensky, P. Nelson, and C. S.
O'Hern, Europhys. Lett. \textbf{38}, 237 (1997); J. F. Marko,
Europhys. Lett. \textbf{38}, 183 (1997).

\bibitem{levinthal56nelson99}  C. Levinthal and H. Crane, Proc. Natl.
Acad. Sci. USA \textbf{42}, 436 (1956);P. Nelson, Proc. Natl.
Acad. Sci. USA \textbf{96}, 14342 (1999).

\bibitem{vankampen}  N. G. van Kampen, \textit{Stochastic Processes in
Physics and Chemistry} (North-Holland, Amsterdam, 1992).

\bibitem{satchwell}  S. C. Satchwell, H. R. Drew, and A. A. Travers,
J. Mol. Biol. \textbf{191}, 659 (1986).

\bibitem{shrader89shrader90}  T. E. Shrader and D. M. Crothers, Proc. Natl.
Acad. Sci. USA \textbf{86}, 7418 (1989); T. E. Shrader and D. M.
Crothers, J. Mol. Biol. \textbf{216}, 69 (1990).

\bibitem{footnote1} For simplicity we address the parameters $q_{k}$ and
$m_{k}$ to the bp rather than dinucleotides.

\bibitem{Cacchione97}  S. Cacchione, M. A. Cerone, and M. Savino,
FEBS Lett. \textbf{400}, 37 (1997).

\bibitem{Mattei02}  S. Mattei, B. Sampaolese, P. De Santis, and M.
Savino, Biophys. Chem. \textbf{97}, 173 (2002); C. Anselmi, G.
Bocchinfuso, P. De Santis, M. Savino, and A. Scipioni, Biophys. J.
\textbf{79}, 601 (2000).

\bibitem{lorch99}  Y. Lorch, M. Zhang, and R. D. Kornberg, Cell
\textbf{96}, 389 (1999).



\end{thebibliography}
\end{document}